\documentclass[a4paper,11pt]{article}
\usepackage{pos}
\usepackage{subcaption}

\title{Progress on the soft anomalous dimension in QCD}
\usepackage{amsmath} 

\usepackage[utf8]{inputenc} 
\usepackage{graphicx}
\usepackage[dvipsnames]{xcolor}
\definecolor{ao(english)}{rgb}{0.0, 0.5, 0.0}
\author*{Einan Gardi}
\author{Zehao Zhu}

\affiliation{Higgs Centre for Theoretical Physics, 
School of Physics and Astronomy, \\
The University of Edinburgh, Edinburgh EH9 3FD, Scotland, UK}

\emailAdd{Einan.Gardi@ed.ac.uk}
\emailAdd{Z.Zhu-37@sms.ed.ac.uk}

\abstract{We review the state-of-the-art knowledge of IR singularities in multileg QCD amplitudes, identifying the key reasons for the remarkable simplicity of the soft anomalous dimension. We then present a novel strategy to compute this quantity using a lightcone expansion of correlators of semi-infinite Wilson lines by the Method of Regions. Recently, this strategy allowed us to determine the three-loop soft anomalous dimension for amplitudes consisting of a single massive coloured particle with any number of massless ones. It opens the way to computing this quantity for amplitudes involving two heavy particles at three loops and potentially going to higher loop orders.}

\FullConference{17th International Symposium on Radiative Corrections: Applications of Quantum Field Theory to Phenomenology (RADCOR2025)\\
5-10 October 2025\\
Puri, India\\}

\begin{document}
\maketitle

\section{Introduction}

Infrared (IR) singularities are an important feature of on-shell gauge-theory amplitudes.
These singularities admit a highly-constrained, universal structure. Their origin in soft and collinear loop-momentum regions guarantees their all-order factorization from the finite hard amplitude. Using renormalization-group equations, factorization can be used to prove that the singularities of any amplitude exponentiate in terms of the so-called soft anomalous dimensions. This universal quantity is therefore of unique importance in the study of amplitudes~\cite{Agarwal:2021ais}.  

In QCD,  IR singularities present one of the key challenges for collider physics. Their cancellation in observables involves a sum over states of different multiplicity, requiring sophisticated subtraction techniques. The singularities also give rise to enhanced logarithmic corrections, which often need to be resummed to achieve precise predictions. 
In any such application the soft anomalous dimension is essential.

In this talk we will review the state-of-the-art knowledge of the soft anomalous dimension in  multileg scattering amplitudes. We will focus on the reasons for the simplicity of this physical quantity, as well as on the strategies for computing it.  In general, one may consider three strategies: extracting the singularities from amplitudes, e.g.~\cite{Henn:2016jdu}, computing it by renormalizing Wilson-line correlators~\cite{Polyakov:1980ca,Korchemsky:1987wg,Aybat:2006mz,Ferroglia:2009ep,Ferroglia:2009ii,Grozin:2014hna,Grozin:2015kna,Almelid:2015jia,Almelid:2016lrq,Bruser:2019auj,Gardi:2025ule,Gardi:2025lws} and determining it using a bootstrap technique~\cite{Almelid:2017qju}, writing down an ansatz in terms of iterated integrals and fixing their rational coefficients using various constraints.
We will primarily focus on the second approach, where significant progress has recently been made, but we will also emphasise insights regarding the properties of the soft anomalous dimension, which will hopefully lead to new results in the future using the third approach.

Recent progress reported here concerns the determination of soft singularities for amplitudes involving both massive quarks and massless quarks and gluons. This progress was achieved thanks to a new approach to evaluate Wilson-line correlators near the lightcone using the Method of Regions (MoR), which was developed and applied in Refs.~\cite{Gardi:2025ule,Gardi:2025lws}.

Let me begin by briefly summarising the state-of-the-art knowledge of IR singularities in multileg amplitudes. For massless scattering, IR singularities in amplitudes with any number of legs are known to three loops since 2015, when the corresponding three-loop soft anomalous dimension was computed~\cite{Almelid:2015jia}.  
The striking simplicity of this three-loop result provided an impetus to try reproducing it using a bootstrap technique~\cite{Almelid:2017qju}.
Considerations pertaining to the general structure of the result, its symmetries and the type of special functions it involves have been studied~\cite{Almelid:2015jia,Almelid:2017qju} allowing a suitable ansatz to be introduced. Then,  
constraints based on collinear and Regge limits have been employed to recover the computed result. This bootstrap programme has been pushed forward to four loops~\cite{Vladimirov:2017ksc,Becher:2019avh,Falcioni:2021buo}, but there is not yet enough information to fully determine the soft anomalous dimension at this order.

Turning now to the massive case, the soft anomalous dimension for amplitudes with an arbitrary number of massive particles is known only to two loops~\cite{Ferroglia:2009ii}. Much progress has been achieved in determining IR singularities for colour-singlet processes involving two massive particles, corresponding to the angle-dependent cusp anomalous dimension~\cite{Korchemsky:1987wg}. This quantity has been known to three loops since 2014~\cite{Grozin:2014hna,Grozin:2015kna}, and for small angles it is also known at four loops~\cite{Grozin:2022wse}. Progress has been slow on the multileg frontier, where integrals depending on several cusp angles prove to be challenging; a step towards addressing these challenges has been taken in Ref.~\cite{Henn:2023pqn}, showing that it is possible to compute the soft anomalous dimension using integrals with fewer IR regulators, reducing the overall complexity.  Nonetheless, the all-massive multileg computation beyond two loops is still beyond reach.  

Another important direction is the determination of the soft anomalous dimension for amplitudes involving both massive and massless particles beyond two loops. In 2022 the soft anomalous dimension for scattering of one massive particle and two massless ones has been computed using a new method by ZeLong Liu and Nicholas Schalch~\cite{Liu:2022elt}. Furthermore, very recently, employing a different approach we completed~\cite{Gardi:2025lws} the computation of the three-loop one-mass soft anomalous dimension, that is the three-loop soft anomalous dimension for amplitudes containing a massive particle and any number of massless ones. This result, and the technique that made its calculation possible, are the main topics of this talk. 

\section{The soft anomalous dimension for massless scattering}

Let me now recall the well-known all-order factorized structure of on-shell \emph{massless} amplitudes~${\cal M}$. We consider fixed-angle scattering of $n$ on-shell massless quarks and gluons, $p_i^2=0$, where all kinematic invariants are taken simultaneously large compared to the QCD scale,  $s_{ij}=2p_i\cdot p_j=2\beta_i\cdot \beta_j Q^2\gg \Lambda^2$.
IR singularities arise from soft and collinear regions of loop momenta, and can be factorized into soft and jet functions, respectively, leaving behind an entirely IR-finite hard amplitude~\cite{Collins_2011,Libby:1978qf,Sen:1982bt,Catani:1998bh,Kidonakis:1998nf,Kidonakis:1997gm,Sterman:2002qn,Gardi:2005yi,Aybat:2006mz,Dixon:2008gr,Gardi:2009qi,Gardi:2009zv,Becher:2009cu,Becher:2009qa,Ma:2019hjq,Feige:2014wja,Almelid:2015jia,Almelid:2016lrq,Almelid:2017qju,Becher:2019avh,Falcioni:2021buo,Maher:2023jqy,Falcioni:2019nxk,Feige:2014wja}:
\begin{equation}
\label{SHJFacorization}
\mathcal{M}_{K}(\{p_i\}, \epsilon)
=
\sum_{N}
S_{KN}( \{ \beta_i \cdot \beta_j \}, \epsilon )\;
H_{N}
\left(
  \left\{s_{ij},{(2 p_i  \cdot n_i)^2}/{n_i^{2}}
  \right\}
\right)
\prod_{i=1}^{n}
\frac{J_i\left((2 p_i \cdot n_i)^2/{n_i^{2}},\,
  \epsilon
\right)}{{\cal J}_i\left((2 \beta_i \cdot n_i)^2/{n_i^{2}},\,
  \epsilon
\right)}
.
\end{equation}
The components in this factorization formula (for a review see Ref.~\cite{Gardi:2009zv})   have gauge-invariant operator definitions in terms of fields and Wilson lines.  
The soft function ${\cal S}$ is defined exclusively in terms of semi-infinite Wilson lines. It depends on all external velocities through their scalar products $\beta_i \cdot \beta_j$. It is a matrix in colour-flow space, mixing between the components of the hard function $H$ in Eq.~(\ref{SHJFacorization}); $K$ and $N$ are indices in this space. Thus, the soft function captures the long-distance interaction correlating between colour and kinematic dependence in the entire process. 
Each partonic jet $J_i$ captures collinear singularities with respect to a given external parton~$i$. It depends on the spin of that parton, but it has trivial colour and kinematic dependence: it is colour singlet and it depends on a single Lorentz invariant $(2 p_i \cdot n_i)^2/{n_i^{2}}$ measuring the momentum of parton $i$ in the direction $n_i$, an auxiliary four vector. Finally, the denominator ${\cal J}_i$ in Eq.~(\ref{SHJFacorization}) is the so-called eikonal jet function. It is defined similarly to the partonic jet $J_i$, except that the parton field is replaced by a Wilson line. Its role in the factorization formula is to remove the double counting owing to the overlap between collinear and soft contributions contained in $J_i$ and $S$, respectively. 

This soft function $S$ is a universal object, describing long-distance interaction in any amplitude. It depends only on the colour and kinematic degrees of freedom of the external particles, but not on their spin, nor does it depend on the details of the hard interaction. This function and its generalization to the case where some of the particles are massive, is the topic of this talk.

IR singularities of scattering amplitudes can be most compactly expressed as an exponential of an integral over the soft anomalous dimension~\cite{Catani:1998bh,Sterman:2002qn,Aybat:2006wq,Aybat:2006mz,Becher:2009cu,Becher:2009qa,Gardi:2009qi,Gardi:2009zv,Almelid:2015jia,Almelid:2017qju,Falcioni:2021buo}:
\begin{equation}
\label{SoftADMassless}
{{\cal M} \left(\frac{p_i}{\mu}, \alpha_s, \epsilon \right)  \, = \, }
{\rm P} \exp\Bigg\{-
\frac12 \int_0^{\mu^2}\frac{d\tau^2}{\tau^2} {\mathbf \Gamma}\left(\tau,\alpha_s(\tau^2,\epsilon)\right)\Bigg\}\,{ {\cal H} \left( \frac{p_i}{\mu},  \alpha_s\right) }\,,
\end{equation}
where the exponential factor contains all IR singularities, both soft and collinear, while the hard amplitude ${\cal H}$, on which it acts, is IR finite. 
The soft anomalous dimension ${\mathbf \Gamma}$ is itself a finite function of the hard-particle momenta and their colour changes, as well as the $4-2\epsilon$-dimensional running coupling and the renormalization scale $\tau$; the singularities are all generated by integrating over $\tau$ from zero momentum.

The soft anomalous dimension is process independent, and it is remarkably simple compared to the finite part of scattering amplitudes. The key reasons for this simplicity are the Soft-Jet-Hard factorization property summarised by Eq.~(\ref{SHJFacorization}) and rescaling-invariance of soft singularities with respect to the momenta (or 4-velocities) of the energetic particles.  
The latter is a property of soft singularities -- not collinear ones. It applies most straightforwardly for soft emission from massive particles, which can be described by timelike Wilson lines, as will be discussed in the next section.

The way rescaling-invariance of soft singularities  manifests itself in the soft anomalous dimension for  massless scattering is complicated by the presence of collinear singularities, which do depend on dimensionful momenta, or energies of the emitters. Thus, in Eq.~(\ref{SHJFacorization}) rescaling invariance holds, strictly speaking, only for the so-called reduced soft function~\cite{Dixon:2008gr,Gardi:2009qi,Gardi:2009zv}: 
\begin{equation}
\label{Reduce_soft_function}
\bar{S}
\left(
\left\{\rho_{ij}\right\}\right)
\equiv \frac{S( \{ \beta_i \cdot \beta_j \}, \epsilon )}{\prod_{i=1}^{n}{\cal J}_i\left((2 \beta_i \cdot n_i)^2/{n_i^{2}},\,  \epsilon\,\right)}\,, \qquad \quad \rho_{ij}\equiv\frac{(\beta_i\cdot \beta_j)^2 n_i^{2}n_j^{2}}
{(2 \beta_i \cdot n_i)^2 (2 \beta_j \cdot n_j)^2} \,.
\end{equation}
This way, rescaling violation in $S$ is fully compensated by the eikonal jets ${\cal J}_i$, and hence the  soft anomalous dimension $\mathbf{\Gamma}$ in Eq.~(\ref{SoftADMassless}) satisfies the following differential equation to all orders~\cite{Becher:2009cu,Becher:2009qa,Gardi:2009qi,Gardi:2009zv},
\begin{equation}
\label{MasslessDiff}
    \sum_{j\neq i}\frac{d{\mathbf{\Gamma}}}{d\log(-s_{ij})} =
    \Gamma^{\text{cusp}}_i(\alpha_s) =
    C_i {\gamma}_{\text{cusp}}(\alpha_s) +\sum_R \frac{d_{RR_i}}{N_{R_i}} g_R(\alpha_s) +\cdots,
\end{equation}
where $\Gamma^{\text{cusp}}_i(\alpha_s)$ is the lightlike cusp anomalous dimension, $C_i$ and $d_{RR_i}/N_{R_i}$ are the quadratic and quartic Casimirs of parton $i$, respectively, while $R$ in the quartic term stands for representations of particles propagating in loops. Here
${\gamma}_{\text{cusp}}(\alpha_s)$ starts at one loop~\cite{Polyakov:1980ca,Korchemsky:1985xu,Korchemsky:1987wg,Korchemsky:1988hd,Korchemsky:1988si}, while $g_R(\alpha_s)$ starts at four loops~\cite{Moch:2004pa,Henn:2016men,Davies:2016jie,Henn:2016wlm,Lee:2017mip,Moch:2017uml,Grozin:2018vdn,Moch:2018wjh,Lee:2019zop,Henn:2019rmi,vonManteuffel:2019wbj,Henn:2019swt,vonManteuffel:2020vjv,Agarwal:2021zft}.
The ellipsis represent higher Casimir contributions. 
The solution of the inhomogeneous differential equation (\ref{MasslessDiff}) induced by the quadratic Casimir term is the well-known dipole formula,
\begin{equation}
\label{Dipole_formula}
{\mathbf \Gamma}_{\text{Dip}}\equiv \sum_{i<j}\mathbf{T}_{i}\cdot\mathbf{T}_{j}\gamma_{\text{cusp}}(\alpha_s)\log\left(-\frac{\tau^2}{2p_i\cdot p_j}\right) \,+\,\sum_{i}\gamma_i(\alpha_s)\,,
\end{equation}
which furnishes the complete IR structure of $\mathbf{\Gamma}$ for massless scattering to two loops. 

One observes that while ${\mathbf \Gamma}$ is a matrix in colour flow space, the right-hand side in Eq.~(\ref{MasslessDiff}) is proportional to the unit matrix. 
The significance of this equation is that it strongly constraints the structure of corrections to the soft anomalous dimension to all orders in perturbation theory. Corrections going beyond the dipole formula fall into two categories: (a) inhomogeneous solutions induced by the quartic (and higher) Casimir terms~\cite{Becher:2009cu,Becher:2009qa,Gardi:2009qi,Gardi:2009zv,Becher:2019avh,Falcioni:2021buo}. These are directly prescribed by Eq.~(\ref{MasslessDiff}) along with multi-loop corrections to the lightlike cusp anomalous dimension;
(b) homogeneous solutions, which depend exclusively on conformally-invariant cross ratios of the energetic particle momenta,
\begin{equation}
    \label{rhoijkl}
\rho_{ijkl}\equiv \frac{(p_i\cdot p_j)(p_k\cdot p_l)}{(p_i\cdot p_k)(p_j\cdot p_l)}\,=\frac{(\beta_i\cdot \beta_j)(\beta_k\cdot \beta_l)}{(\beta_i\cdot \beta_k)(\beta_j\cdot \beta_l)}\,.
\end{equation}
Therefore, the soft anomalous dimension for massless scattering takes the form
\begin{align}
  \mathbf{\Gamma}= {\mathbf \Gamma}_{\text{Dip}}+\mathbf{\Delta}\left(\left\{\rho_{ijkl}\right\}\right) +{\cal O}(\alpha_s^4)\,,
\end{align}
where we suppressed four loop corrections induced by the quartic Casimir contributions to the cusp anomalous dimension in Eq.~(\ref{MasslessDiff}). We refer the interested reader to  Refs.~\cite{Becher:2019avh,Falcioni:2021buo,Duhr:2025cye} for further details regarding four-loop contributions.

The three-loop soft anomalous dimension was computed in Ref.~\cite{Almelid:2015jia,Almelid:2016lrq} (see also Ref.~\cite{Almelid:2017qju}) and it takes the form:
\begin{equation}
\label{Delta_massless}
\mathbf{\Delta}\left(\left\{\rho_{ijkl}\right\}\right)=    
\sum_{i<j<k<q}\sum_{(u,v;w)\in P} \mathbf{T}_{uv;w q}
 {\cal F}_{0,4}(\rho_{uwv q},\rho_{vwu q }) 
+
{\cal F}_{0,3}\, \sum_{i}\sum_{j<k;\,j,k\neq i}\mathbf{T}_{ij;ik}
\end{equation}
where 
\begin{equation}
{\cal F}_{h,l}=\,\,\sum_{n=3}^{\infty}\left(\frac{\alpha_s}{4\pi}\right)^n{\cal F}_{h,l}^{(n)}
\end{equation}
and where $(u,v;w)\in P\equiv \{(i,k;j),(j,k;i),(i,j;k)\}$ defines the three quadrupole channels, having colour structures 
\begin{equation}
\label{Tuvwq}
\mathbf{T}_{uv;w q}\equiv
f^{abe}f^{cde}
\{\mathbf{T}_{u}^a,
\mathbf{T}_{v}^b,
\mathbf{T}_{w}^c,
\mathbf{T}_{q}^d
\}_{+}\,,
\end{equation}
where the curly brackets are defined, as in~Refs.~\cite{Becher:2019avh,Falcioni:2021buo,Duhr:2025cye}, as a symmetric sum over all permutations of the $n$ generators it contains, divided by~$n!$. 
For a given set of four partons $(i,j,k,l)$ this function depends on the kinematics through two independent rescaling-invariant cross ratios. These can be conveniently traded for the variables $z$ and $\bar{z}$:
\begin{align}
\label{zzbVar}
\begin{split}
\rho_{ijkq}\equiv\,\,&z\bar{z},
\qquad
\rho_{kjiq}\equiv(1-z)(1-\bar{z})\,.
\end{split}
\end{align} 
The explicit result reads
\begin{eqnarray}
 \label{Fmassless}
  {\cal F}_{0,4}^{(3)}(\rho_{ijkq},\rho_{kjiq})&=&\,\,16\left[F_{0,4}(1-\bar{z},1-z)-F_{0,4}(z,\bar{z})\right],
 \end{eqnarray}
 with
 \begin{eqnarray}
F_{0,4}(z,\bar{z})\equiv\,\,\mathcal{L}_{10101}(z)+\,2\zeta(2)\left[\mathcal{L}_{100}(z)+\mathcal{L}_{001}(z)\right]\,, 
\end{eqnarray}
and 
\begin{equation}
 \label{F03massless}
    {\cal F}_{0,3}^{(3)}=\,\,32\left[2\zeta(2)\zeta(3)+\zeta(5)\right]\equiv 32C,
\end{equation}
where $\mathcal{L}(z)$ are single-valued~\cite{Brown:2004ugm,Dixon:2012yy} harmonic polylogarithms. They are single valued when $z=(\bar{z})^*$. The expansion of $F_{0,4}(z,\bar{z})$ in ordinary (multi-valued) Multiple Polylogarithms (MPLs) of $z$ and $\bar{z}$ has been provided in Ref.~\cite{Almelid:2015jia}. Note that ${\cal F}_{0,4}$ in Eq.~(\ref{Fmassless}) is antisymmetric under $z\leftrightarrow 1-\bar{z}$, in line with Bose symmetry and the antisymmetry of the colour structure $\mathbf{T}_{ik;j q}$ under $i\leftrightarrow k$ permutation.

\section{Correlators of timelike Wilson lines}


In the previous section we reviewed what is known about the soft anomalous dimension for massless scattering. 
At this point we wish to recall how these functions are defined and computed. The key observation, which allows us to understand the universal nature and the properties of soft singularities, and ultimately compute them, is that soft gluons couple only to external coloured particles participating in the hard process. This coupling is an Eikonal coupling, that is, it is independent of the spin and energy of the emitter. It is the same as coupling to a classical source provided by Wilson lines that follow the classical trajectories of the energetic particles, and carrying the same colour charge. 

The soft singularities can therefore be computed by evaluating Wilson-line correlators.  Let us consider now $n$ semi-infinite Wilson lines, each carrying colour      in an arbitrary representation of the colour group, extending from the hard interaction vertex to infinity in the directions $\beta_l$, for $l=1,2,\ldots, n$, where $\beta_l^2>0$. We assume that colour is conserved at the hard interaction vertex and define
\begin{equation}
\label{calSdef}
{\cal S}= \left<\Phi_{\beta_1} \otimes  \Phi_{\beta_2} \otimes \cdots \otimes\Phi_{\beta_n}\right>\,,\qquad \Phi_{\beta_l}={\cal P} \exp\left[ig_s\int_{0}^{\infty} dt\beta_l\cdot A(t\beta_l)\right]\,.
\end{equation}
The universality and other key properties of the soft function stem directly from this definition.
Specifically, one can deduce that ${\cal S}$ admits the following properties:
\begin{itemize}
\item{} \emph{diagrammatic exponentiation} in terms of connected colour factors~\cite{
Frenkel:1984pz,Gatheral:1983cz,Sterman:1981jc,Mitov:2010rp,Gardi:2010rn,Gardi:2011wa,Gardi:2011yz,Gardi:2013ita,Agarwal:2020nyc,Agarwal:2022xec,Mishra:2023acr,Agarwal:2024srg}\,.
\item{} \emph{rescaling invariance} with respect to any of the Wilson-line velocities, implying that the function ${\cal S}$ depends on the kinematics only through the scalar product of normalized velocities, namely \begin{align}
\label{cuspangle}
    -\alpha_{IJ}-\frac{1}{\alpha_{IJ}}\equiv 
    \gamma_{IJ}\equiv \frac{2\beta_I\cdot\beta_J}{\sqrt{\beta_I^2}\sqrt{\beta_J^2}}\equiv v_I\cdot v_J.
\end{align}
\item{} \emph{multiplicative renormalizability} of the vertex where Wilson lines meet~\cite{Brandt:1981kf}. This property is a generalization of the renormalization of the cusp singularity of a Wilson loop~\cite{Polyakov:1980ca,Arefeva:1980zd,Dotsenko:1979wb,Korchemsky:1985xj,Korchemsky:1987wg}. 
\end{itemize}
We have considered here timelike Wilson lines, not only because we wish to generalise the discussion of soft singularities to the massive case. The requirement that the lines are non-lightlike is essential for the rescaling symmetry of ${\cal S}$ as well as for the multiplicative renormalizability of this correlator. If any of the Wilson lines is taken to be strictly lightlike, $\beta_l^2=0$, then these two properties break down due to collinear singularities. 
One therefore sets up the computation of the soft function using a correlator of  timelike lines, $\beta_l^2>0$ even if one is ultimately interested in the massless limit, $\beta_l^2\sim p_l^2\to 0$.

Correlators of  semi-infinite Wilson lines give rise to scaleless integrals, which vanish in dimensional regularization. This embodies the fact that IR singularities (which can be identified with those of the partonic amplitude) are mirrored by UV singularities associated with the hard vertex where the Wilson lines meet. The scaleless integrals one obtains upon evaluating Eq.~(\ref{calSdef}) feature both IR and UV divergences.    
Upon regularizing the correlator in the IR (see e.g.~\cite{Korchemsky:1987wg,Gardi:2011yz,Almelid:2015jia,Henn:2023pqn}), multiplicative renormalizability of the hard vertex allows us to compute the soft anomalous dimension~$\mathbf{\Gamma}$.
A convenient way of regularizing the correlator in the IR is to modify the definition of the Wilson line such that
\begin{align}
\label{regWilson}
    \Phi^{(m)}_\beta\equiv {\cal P}\exp\left[ig_s\int_0^\infty dt e^{-imt\sqrt{\beta^2-i\varepsilon}}\beta\cdot A(t\beta)\right],\qquad m>0,
\end{align}
where an exponential regulator is used to suppress the long-distance radiation. Therefore, only the UV  singularities of the correlator 
\begin{align}
    \left<\Phi^{(m)}_{\beta_1}  \Phi^{(m)}_{\beta_2} \cdots\Phi^{(m)}_{\beta_n}\right>
\end{align}
are captured by dimensional regularization. These  correspond to the IR singularities of the amplitudes. Because of the multiplicative renormalizability, such UV divergences may be removed by multiplying the correlator by a $Z$ factor, 
\begin{align}
    \left<\Phi^{(m)}_{\beta_1}   \Phi^{(m)}_{\beta_2}  \cdots \Phi^{(m)}_{\beta_n}\right>_{\text{ren}}\equiv \left<\Phi^{(m)}_{\beta_1} \Phi^{(m)}_{\beta_2}  \cdots \Phi^{(m)}_{\beta_n}\right>{\mathbf Z}_{\text{UV}}(\alpha_s(\mu)),
\end{align}
where $\mu$ is the renormalization point and the subscript ``$\text{ren}$'' indicates that the renormalized correlator is finite. The anomalous dimension of the factor ${\mathbf Z}_{\text{UV}}$ is the soft anomalous dimension of  amplitudes,
\begin{align}
    \frac{d}{d\log\mu}{\mathbf Z}_{\text{UV}}=-{\mathbf Z}_{\text{UV}}{\mathbf \Gamma}_{\cal S}.
\end{align}
The soft anomalous dimension ${\mathbf \Gamma}_{\cal S}$ is the same as ${\mathbf \Gamma}$ in Eq.~\eqref{SoftADMassless}, but for the case where all the particles are massive (where collinear singularities do not appear).
This is the standard method of computing this quantity.

\begin{figure}[t]
\centering
\setlength{\tabcolsep}{10pt} 
\renewcommand{\arraystretch}{1.2} 

\begin{tabular}{ccc}
\begin{subfigure}[t]{0.25\textwidth}
  \centering
  \includegraphics[width=\linewidth]{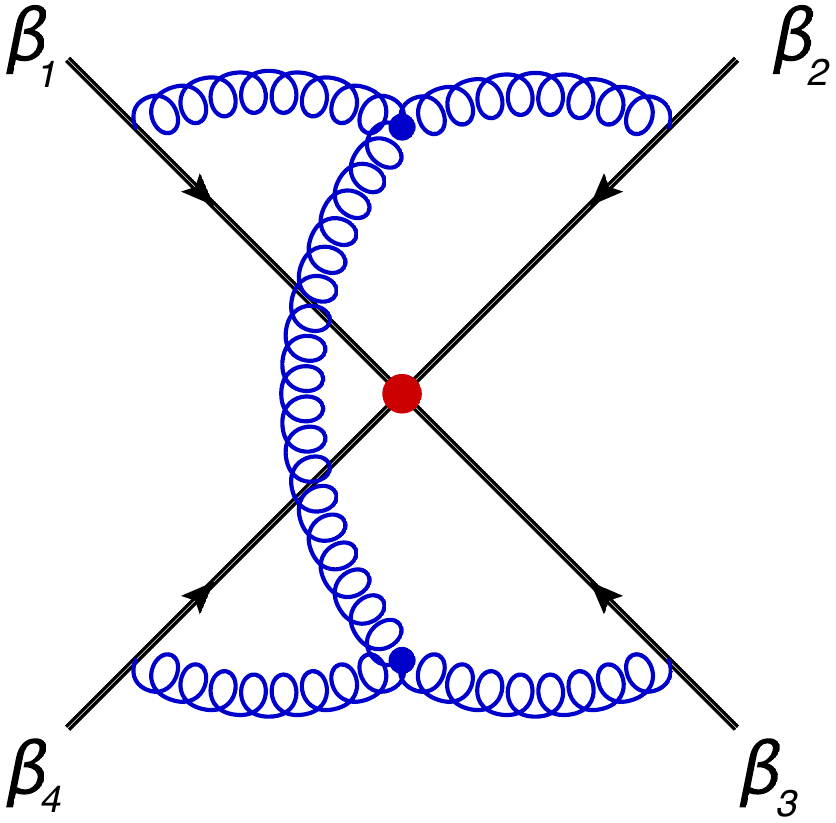}
  \caption*{(a) 4 massless particles} 
\end{subfigure}
&
\begin{subfigure}[t]{0.25\textwidth}
  \centering
  \includegraphics[width=\linewidth]{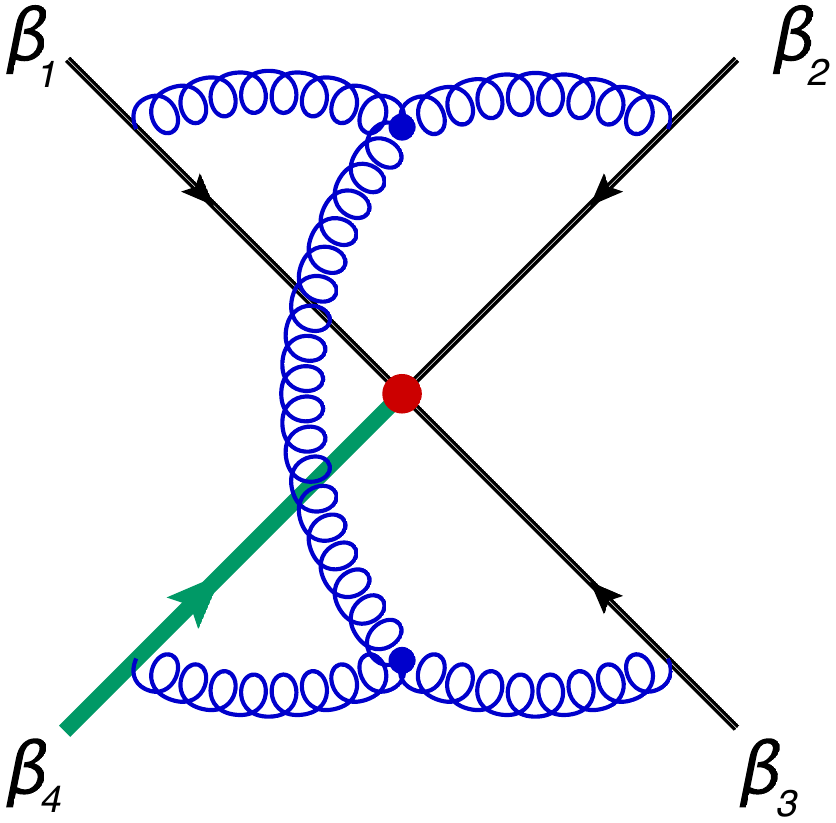}
  \caption*{(b) 1 massive and 3 massless}
\end{subfigure}
&
\begin{subfigure}[t]{0.25\textwidth}
  \centering
  \includegraphics[width=\linewidth]{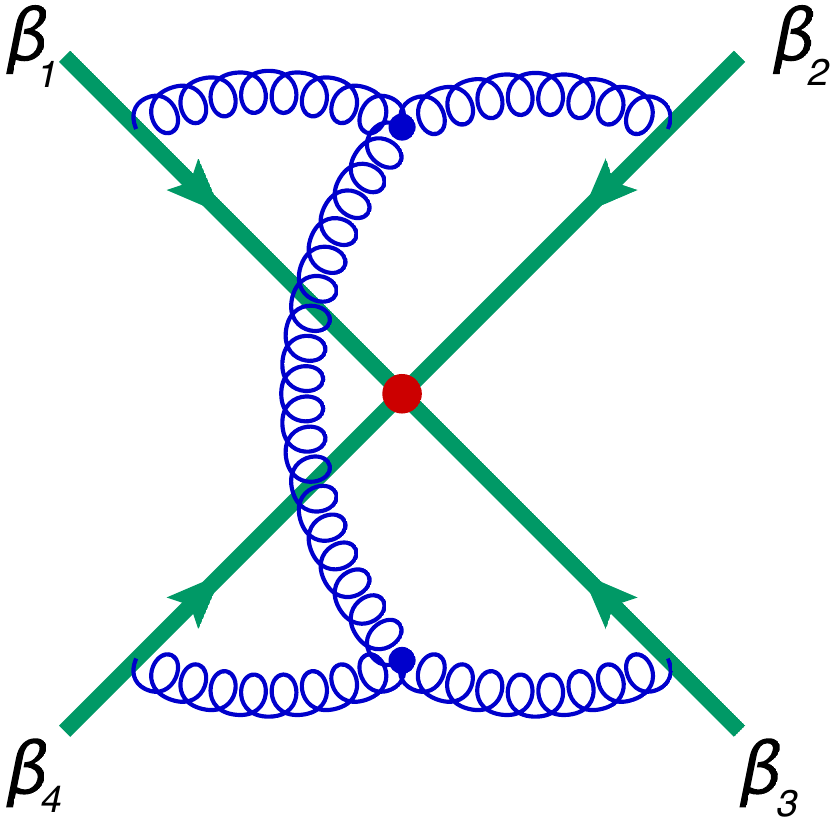}
  \caption*{(c) 4 massive particles}
\end{subfigure}
\\
\\
$\displaystyle
\rho_{ijkl}\equiv\frac{(\beta_i\cdot \beta_j)(\beta_l\cdot \beta_l)}{(\beta_i\cdot \beta_k)(\beta_j\cdot \beta_l)}
$ &
$\displaystyle
r_{ijQ}\equiv\,\,
\frac{\beta_Q^2\beta_i\cdot\beta_j}{2\beta_i\cdot\beta_Q \beta_j\cdot\beta_Q}
$ &
$\displaystyle
\gamma_{IJ}\equiv \frac{2\beta_I\cdot\beta_J}{\sqrt{\beta_I^2}\sqrt{\beta_J^2}}
$
\\

$\displaystyle
\{\rho_{1234}, \rho_{3214}\}
$ &
$\displaystyle
\{r_{12\color{ao(english)}{4}}, r_{23\color{ao(english)}{4}}, r_{13\color{ao(english)}{4}}\}

$ &
$\displaystyle
\{{\color{ao(english)}{ \gamma_{12}, \gamma_{13}, \gamma_{14}, \gamma_{23}, \gamma_{24}, \gamma_{34}}}\}
$
\end{tabular}

\caption{Three different scattering  configurations involving massless and massive coloured particles, along with the corresponding sets of independent rescaling-invariant  kinematic variables they depend upon.}
\label{fig:number_of_Kin_var}
\end{figure}

As emphasised above, when computing soft singularities using Wilson-line correlators one begins with timelike lines (so as to ensure  multiplicative renormalizability) even if the interest is in the lightlike limit. Yet, it is clear that the lightlike limit offers a tremendous simplification compared to the generic timelike case. In order to illustrate that, we show in Fig.~\ref{fig:number_of_Kin_var} the number of independent rescaling-invariant kinematic variables in different configurations, after the lightlike limit has been taken for some of the Wilson lines. In the all-massless case (in Fig.~\ref{fig:number_of_Kin_var} this is case~(a)), corresponding to the result presented in Eq.~(\ref{Delta_massless}), there are just \emph{two} independent cross ratios that can be formed out of a set of four lightlike velocities; the result for the soft anomalous dimension is indeed remarkably simple. In the other extreme there is the all-massive case (this is case (c)) where \emph{six} independent cusp angles~(\ref{cuspangle}) can be formed of any chosen pair out of the four timelike velocities. The corresponding three-loop computation is difficult and remains beyond the state of the art. There are three intermediate cases, where the number of massive particles gradually increases -- the simplest of these is the one-mass case, shown in (b), where one Wilson line is timelike and the three others are lightlike. Here there are \emph{three} independent rescaling-invariant ratios that can be formed from the four velocities (Eq.~(\ref{newVar}) below). The corresponding soft anomalous dimension, involving one massive particle with three massless ones, is the one we recently completed~\cite{Gardi:2025lws,GZ-TBP} and will discuss in what follows.

\section{New strategy for computing the soft anomalous dimension}

Our concluding discussion in the previous section  focused on the number of kinematic variable as a measure of the complexity of the soft anomalous dimension function. We have seen that taking the lightlike limit translates into a major simplification. 
However, we have not yet addressed the question \emph{how to compute this  limit} when it is clear that the all-massive integral is itself too difficult to evaluate by present methods.
In other words, one needs to compute the leading term in the asymptotic expansion of the relevant integrals in the limit where certain Wilson lines become lightlike. Importantly, the limit cannot be taken prior to integration, as the two do not commute. There are several ways to approach this problem. 

The strategy used Ref.~\cite{Almelid:2015jia,Almelid:2016lrq,Gardi:2016ttq} was based on first integrating the web diagrams so as to explicitly isolate  the coefficients of the $1/\epsilon$ UV pole, which  have been expressed as multi-dimensional Mellin-Barnes integrals. At the next stage these Laurent coefficients have been expanded in the limit were all four Wilson lines simultaneously become lightlike. In this expansion a major simplification occurs, allowing one to express the integrals as more compact iterated sums, which can be evaluated in terms of~MPLs. The result for individual integrals still depends of all six variables, albeit four of which appear only through (powers of) logarithms. Ultimately, the anomalous dimension becomes a simple function of two independent cross ratios, Eq.~(\ref{Fmassless}) above. The computation is challenging, and it would be hard to repeat in a less extreme limit in which one or more of the lines remain timelike. 

The new strategy proposed and applied in Refs.~\cite{Gardi:2025lws,Gardi:2025ule} relies instead on the Method of Regions (MoR)~\cite{Beneke:1997zp,Smirnov:2002pj,Pak:2010pt,Jantzen:2011nz,Jantzen:2012mw,Semenova:2018cwy,Ananthanarayan:2018tog,Heinrich:2023til,Borowka:2017idc,Gardi:2022khw,Ma:2023hrt,Chen:2024xwt}. Its advantage is that by performing the expansion (in each region) first, the integrals we end up computing strictly depend only  on the variables that remain finite in the limit of interest, hence maximizing the simplification offered by that limit. In particular, in the one-mass case~\cite{Gardi:2025ule} the most complicated integrals depend on the three variables 
    \begin{align}
\label{newVar}
\begin{split} r_{ijQ}\equiv\,\,
\frac{\beta_Q^2\beta_i\cdot\beta_j}{2\beta_i\cdot\beta_Q \beta_j\cdot\beta_Q},
\quad\quad
    r_{jkQ}\equiv\,\,
\frac{\beta_Q^2\beta_j\cdot\beta_k}{2\beta_j\cdot\beta_Q \beta_k\cdot\beta_Q},
    \quad\quad
    r_{ikQ}\equiv\,\,
\frac{\beta_Q^2\beta_i\cdot\beta_k}{2\beta_i\cdot\beta_Q \beta_k\cdot\beta_Q},
\end{split}
\end{align}
rather than the six cusp angles. In the rest of this section we briefly describe the calculation method. The result will be discussed in the next section.

We start by considering the correlator of four timelike Wilson lines ($\beta_U^2>0$), 
\begin{align}
\label{def_correlator}\left<\Phi_{\beta_I}^{(m)}\Phi_{\beta_J}^{(m)}\Phi_{\beta_K}^{(m)}\Phi_{\beta_Q}^{(m)}\right>\, =\exp\left[\sum_{n}\left(\frac{\alpha_s}{4\pi}\right)^n w^{(n)}\right],
\end{align}
which is multiplicatively renormalizable.  The Wilson lines are regularised as in  Eq.~\eqref{regWilson}.
We stress that the four velocities are independent: there may be additional particles emanating from the hard vertex where the four lines meet.

Next we use the MoR to perform an asymptotic lightcone expansion in  three of the four squared velocities,
\begin{align}
\label{virtualities_expansion}
    \beta_I^2\sim \lambda_i\,,
    \quad\quad
     \beta_J^2\sim \lambda_j\,,
     \quad\quad
      \beta_K^2\sim \lambda_k\,.
\end{align}
The naive limit -- dubbed the ``hard region'' -- corresponding to taking the squared velocities in Eq.~(\ref{virtualities_expansion}) to zero in the integrand, involves losing IR regularization along the corresponding lightlike lines, therefore generating collinear singularities in addition to the UV ones. This leads to higher-order poles in $\epsilon$, which cannot be linked with the renormalization of the multi-Wilson-line vertex.
However, the full asymptotic expansion obtained by the MoR restores the multiplicatively-renormalizable correlator~\cite{Gardi:2025ule}, replacing the higher-order~$\epsilon$ poles in each web with logarithms of the expansion parameters in Eq.~(\ref{virtualities_expansion}). These logarithms ultimately cancel in the correlator in the sum of all webs, where one can read off the soft anomalous dimension.  

Importantly, in the present application of the MoR there is an algorithmic method to determine the complete set of regions. The key observation is that there exists a Euclidean regime where $\beta_{U}\cdot\beta_{V}<0$ for any pair of velocities $U,V\in\{I,J,K,Q\}$.
Upon performing the asymptotic expansion in this  kinematic regime, we guarantee that all the monomials in the Symanzik graph polynomials in any integral are non-negative, and hence there are no hidden regions~\cite{Gardi:2024axt,Ma:2025emu}. Thus, the complete set of regions can be determined using the facets of the Newton polytope of the graph polynomials~\cite{Pak:2010pt,Jantzen:2012mw,Semenova:2018cwy,Gardi:2022khw,Ananthanarayan:2018tog,Heinrich:2023til,Borowka:2017idc,Chen:2024xwt}. In this work we used the packages {\tt pySecDec}~\cite{Heinrich:2021dbf} and {\tt AmpRed}~\cite{Chen:2024xwt} to determine the region vectors, which correspond to the scaling of all integration variables in any given region. 

\begin{figure*}
\begin{subfigure}{0.18\linewidth}
    \centering \includegraphics[width=1\linewidth]{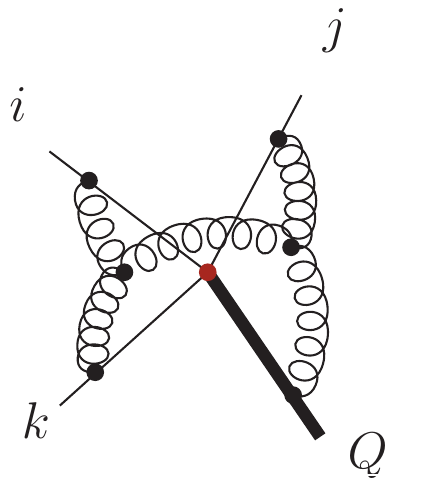}
    \label{fig:11113g3g}
    \end{subfigure}
    \hfill
    \begin{subfigure}{0.18\linewidth}
    \centering \includegraphics[width=1\linewidth]{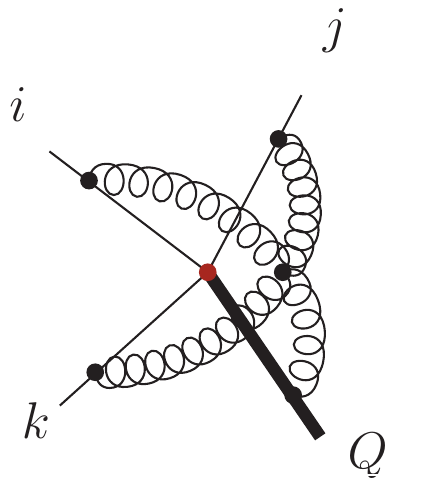}
    \label{fig:11114g}
    \end{subfigure}
    \hfill
     \begin{subfigure}{0.18\linewidth}
    \centering \includegraphics[width=1\linewidth]{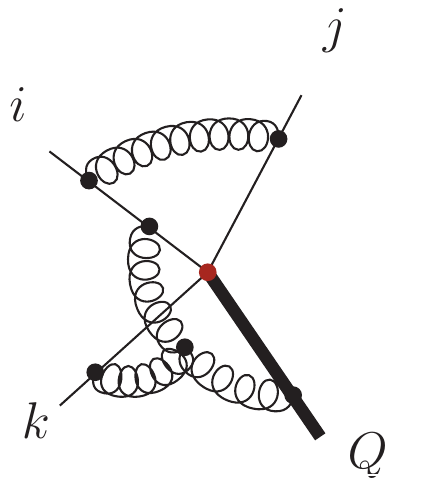}
    \label{fig:1112}
    \end{subfigure}
    \hfill
     \begin{subfigure}{0.18\linewidth}
    \centering \includegraphics[width=1\linewidth]{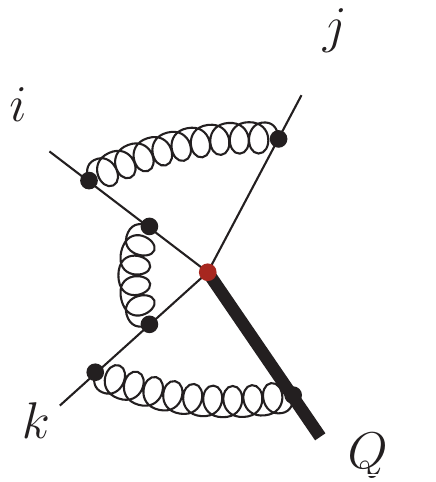}
    \label{fig:1122}
    \end{subfigure}
    \hfill
     \begin{subfigure}{0.18\linewidth}
    \centering \includegraphics[width=1\linewidth]{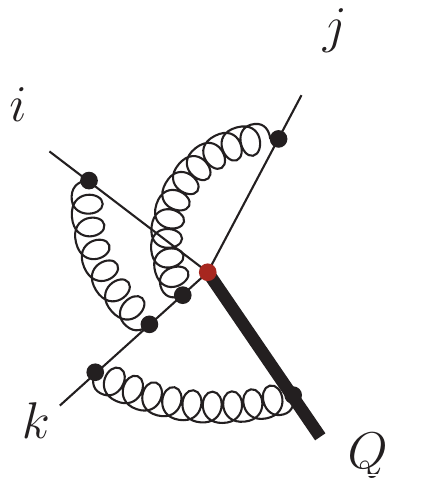}
    \label{fig:1113}
    \end{subfigure}
     \caption{Representative three-loop diagrams contributing to the quadrupole structure. The thick line corresponds to the timelike Wilson line with velocity $\beta_Q$, while the thin lines to Wilson lines with   nearly lightlike velocities. The two leftmost diagrams depict connected webs, $W_{1111}$, the middle one is a representative diagram of the $W_{1112}$ web and the two rightmost ones belong to  multiple gluon-exchange webs,~$W_{1122}$ and  $W_{1113}$, respectively.}
\label{fig:threeloop4linewebs}
\end{figure*}

Performing the perturbative expansion of the correlator~(\ref{def_correlator}), one finds that there are four types of four-line webs (see Fig.~\ref{fig:threeloop4linewebs})~\cite{Frenkel:1984pz,Gatheral:1983cz,Sterman:1981jc,Mitov:2010rp,Gardi:2010rn,Gardi:2011wa,Gardi:2011yz,Gardi:2013ita,Almelid:2015jia,Gardi:2016ttq,Almelid:2016lrq} that contribute to the colour quadrupole structure $\mathbf{T}_{uv;w Q}$ of Eq.~(\ref{Tuvwq}) at three loops~\cite{Almelid:2015jia},
\begin{align}
\label{FourLineWebs}
\left(\frac{\alpha_s}{4\pi}\right)^3w^{(3)}= W_{1111}+W_{1112}+W_{1122}+W_{1113}+\cdots\,,
\end{align}
where the subscripts on $W$ count the number of attachments to each of the four Wilson lines. Each $W$ term  represents the sum of all webs of a given type in the Feynman gauge. The ellipsis represent webs that span fewer lines.
$W_{1111}$ represents the sum of all connected diagrams spanning four lines. These feature either a single 4-gluon vertex or two 3-gluon vertices (the two leftmost diagrams in Fig.~\ref{fig:threeloop4linewebs}).
These diagrams are special: they are the only ones that give rise to integrals depending on all three rescaling-invariant ratios of Eq.~(\ref{newVar}). We therefore focus our description of the calculation on these webs. The structure of $W_{1111}$ is
\begin{eqnarray}
\label{W1111calY}
W_{1111}\,&=&\,\textbf{T}_I^a\textbf{T}_J^b\textbf{T}_K^c\textbf{T}_Q^d\big[f^{abe}f^{cde}{\cal Y}_{(IJ;KQ)}(\{\alpha_{VU}\})
+f^{ace}f^{bde}{\cal Y}_{(IK;JQ)} (\{\alpha_{VU}\})
\\&& \hspace{40mm}
+f^{ade}f^{bce}{\cal Y}_{(JK;IQ)}(\{\alpha_{VU}\})\big]\,.\nonumber
\end{eqnarray}
The three channels are related by permutations, so let us now consider ${\cal Y}_{(IJ;KQ)}$.  Relating the correlator~(\ref{def_correlator}) to the anomalous dimension requires to compute ${\cal Y}$ with four timelike Wilson lines, so prior to the expansion, the kinematic dependence is via the set $\{\alpha_{VU}\}$ of \emph{six} rescaling-invariant variables (\ref{cuspangle}). 
Being connected, these diagrams features a single UV pole in $\epsilon$,
\begin{align}
\label{Laurent}
    {\cal Y}_{(IJ;KQ)}(\{\alpha_{VU}\})=\sum_{l=-1}^{\infty}{\cal Y}_{(IJ;KQ)}^{(l)}(\{\alpha_{VU}\})\,\epsilon^l\,.
\end{align}
The connected nature of the diagrams also guarantees that they have a single overall divergence. As a consequence it is sufficient to regularize the IR on a single Wilson line~\cite{Henn:2023pqn}. We therefore drop the regulator $m$ on the lines $I$, $J$ and~$K$,  while keeping it on line $Q$, without affecting the UV $1/\epsilon$ pole we are interested in. 
The asymptotic expansion ${\cal T}$ of the Laurent coefficients in Eq.~(\ref{Laurent}) takes the form a log-power expansion. Specifically, at the leading order in~$\epsilon$, ${\cal O}(\epsilon^{-1})$, we have
\begin{align}
\label{CalYExpected}
\begin{split}
   &{\cal T}_{\lambda_i,\lambda_j,\lambda_k}\left[ {\cal Y}_{(IJ;KQ)}^{(-1)}(\{\alpha_{VU}\})\right]= 
{\cal Y}_{(ij;kQ)}^{(-1)}(r_{ikQ},r_{jkQ},r_{ijQ}; \lambda_i, \lambda_j, \lambda_k) 
   \,+\, {\cal O}(\lambda)\,,
   \end{split}
\end{align}
where ${\cal Y}_{(ij;kQ)}^{(-1)}$ on the right-hand side 
depends on the kinematics via the ratios 
$r_{uvQ}$ as well as the three logarithms of the expansion parameters $\lambda_u$ of Eq.~(\ref{virtualities_expansion}).

We compute the asymptotic lightcone expansion using the MoR, which 
yields a sum over region integrals:
\begin{eqnarray}
\label{CalYExp}
 & {\cal T}_{\lambda_i,\lambda_j,\lambda_k}\left[ {\cal Y}_{(IJ;KQ)}(\{\alpha_{VU}\})\right]= \displaystyle{\sum_{[n_i,n_j,n_k]}}
\lambda_i^{n_i\epsilon}
\lambda_j^{n_j\epsilon}
\lambda_k^{n_k\epsilon} 
\,\, {\cal Y}_{(ij;kQ)}^{[n_i,n_j,n_k]}(r_{ikQ},r_{jkQ},r_{ijQ})\,,
\end{eqnarray}
where
we neglected all subleading powers in the expansion parameters~$\lambda_i$, $\lambda_j$ and $\lambda_k$. Here the regions are designated by $[n_i,n_j,n_k]$, identifying their scaling law~$\lambda_u^{n_u\epsilon}$ for $u\in \{i,j,k\}$, where $n_u$ are integers. 
In each region integral we have scaled out the non-analytic dependence on the expansion parameters. Crucially, any given region integral ${\cal Y}_{(ij;kQ)}^{[n_i,n_j,n_k]}$~depends \emph{only} on the three $r_{uvQ}$ ratios, and not on the expansion parameters. Individual region integrals feature higher-order poles in $\epsilon$. These cancel in the sum of regions, leaving a trace in ${\cal Y}^{(-1)}_{(ij;kQ)}$ as  logarithms of the three expansion parameters, in line with Eq.~(\ref{CalYExpected}).  

Applying the MoR to $W_{1111}$, we get $56$ regions for each of the three channels. Most of the region integrals are simple to evaluate; the most complicated eight region integrals depend on all three $r_{uvQ}$ ratios of Eq.~(\ref{newVar}) via MPLs. 
These are the hard region, where the three Wilson lines are taken to be strictly lightlike, and seven other regions which feature up to three collinear modes:
\begin{align}
\label{MoRWeb111}
\begin{split}
   {\cal T}_{\lambda_i,\lambda_j,\lambda_k}\left[W_{1111}\right]=&\vcenter{\hbox{\includegraphics[width=1.6cm]{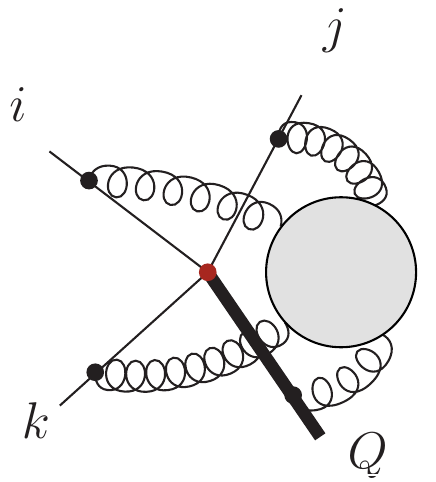}}}=\vcenter{\hbox{\includegraphics[width=1.6cm]{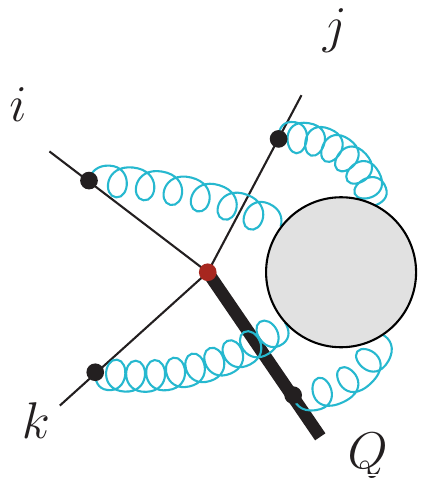}}}+\lambda_i^{-\epsilon}\vcenter{\hbox{\includegraphics[width=1.6cm]{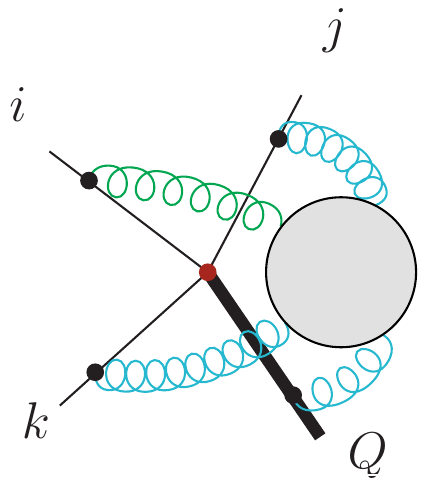}}}+\lambda_j^{-\epsilon}\vcenter{\hbox{\includegraphics[width=1.6cm]{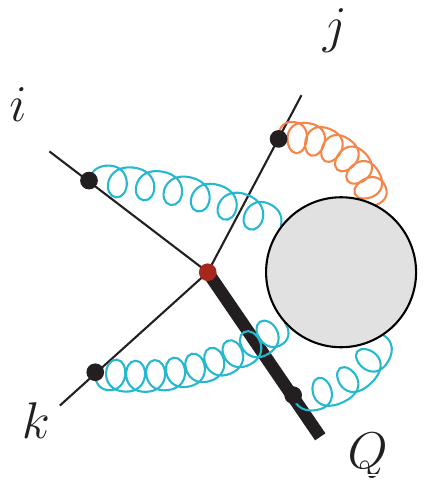}}}+\lambda_k^{-\epsilon}\vcenter{\hbox{\includegraphics[width=1.6cm]{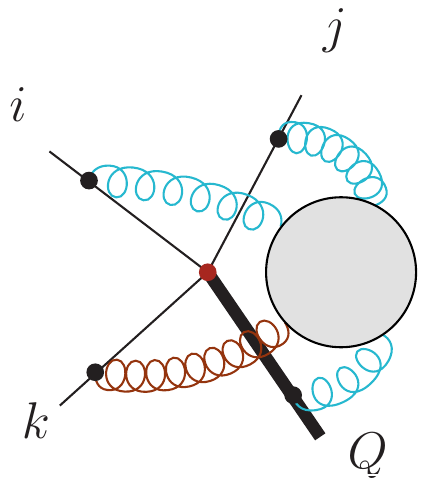}}}
\\&\hspace{-20mm}+\lambda_i^{-\epsilon}\lambda_j^{-\epsilon}\vcenter{\hbox{\includegraphics[width=1.6cm]{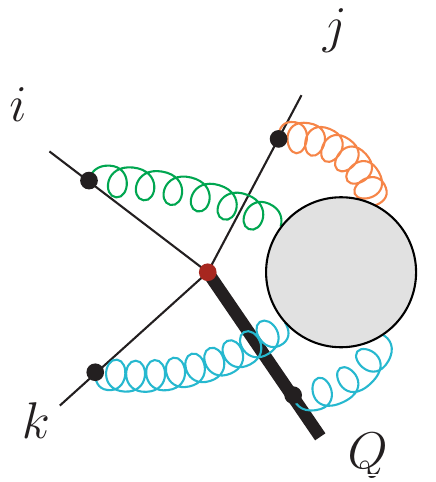}}}+\lambda_i^{-\epsilon}\lambda_k^{-\epsilon}\vcenter{\hbox{\includegraphics[width=1.6cm]{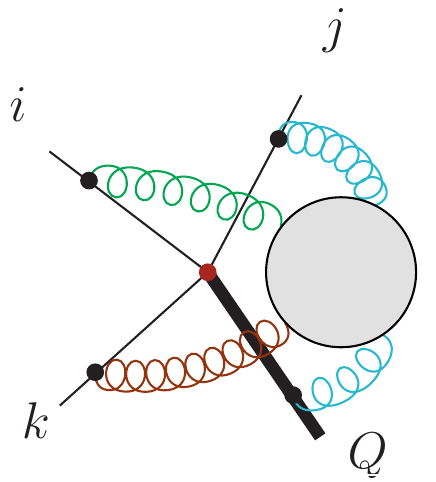}}}+\lambda_j^{-\epsilon}\lambda_k^{-\epsilon}\vcenter{\hbox{\includegraphics[width=1.6cm]{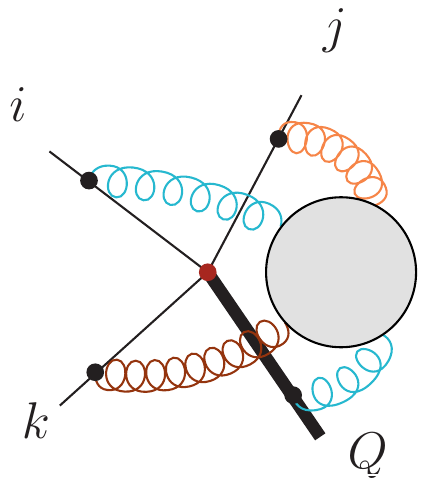}}}+\lambda_i^{-\epsilon}\lambda_j^{-\epsilon}\lambda_k^{-\epsilon}\vcenter{\hbox{\includegraphics[width=1.6cm]{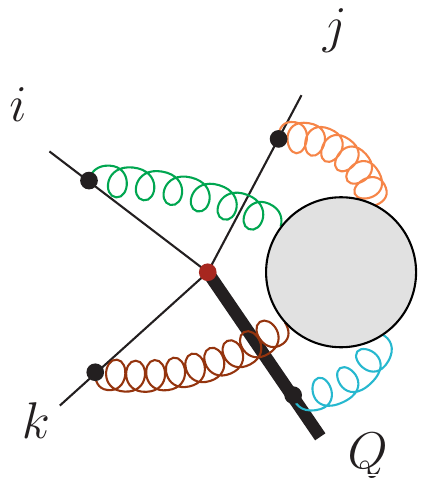}}}
\\&\hspace{-20mm}+\ldots\,.
 \end{split}
 \end{align}
Here the gluons are colour-coded according to the loop-momentum modes~\cite{Gardi:2025ule}: the blue gluons are hard, while the green, orange and brown gluons are collinear to the $i$, $j$ and $k$ directions, respectively.  The blob represents all the possible ways in which the four gluons connect through a single 4-gluon vertex or two 3-gluon vertices.

We use the method of differential equation to compute the region integrals.  Using the packages {\tt AmpRed}~\cite{Chen:2024xwt}, {\tt Kira}~\cite{Maierhofer:2017gsa,Klappert:2020nbg,Lange:2025fba} and {\tt LiteRed}~\cite{Lee:2012cn,Lee:2013mka}, we derive the differential equation for a subset of the regions of one channel, obtaining the rest by permutations. In addition to the singularities at $r_{uvQ}\rightarrow \{0,1,\infty\}$ (represented by the physical alphabet in Eq.~(\ref{alphabet}) below), the differential equations display singularities also at~${\textcolor{blue}{\Delta_1}}\rightarrow 0$ and ${\textcolor{blue}{\Delta_2}}\rightarrow 0$, where
 \begin{align}
 \label{NewSing}
 \begin{split}
     \textcolor{blue}{\Delta_1}=\,\kappa(r_{ijQ},r_{ikQ},r_{jkQ}),
\qquad    \textcolor{blue}{\Delta_2}=\,\,\textcolor{blue}{\Delta_1}+4r_{ijQ}r_{jkQ}r_{ikQ}\,, 
\end{split}
 \end{align}
where $\kappa$ is the K\"{a}ll\'{e}n function, $\kappa(x,y,z) = x^2 + y^2 + z^2 - 2xy - 2yz - 2zx$. $\textcolor{blue}{\Delta_1}$ and $\textcolor{blue}{\Delta_2}$ are quadratic and cubic polynomias in the $r_{ijQ}$ variables, respectively. 
These complicated letters only appear in $W_{1111}$, not in any other contribution to the correlator.   

In order to solve the differential equation, we bring it to an $\epsilon$-factorized form~\cite{Henn:2013pwa}. To this end we first identify one uniformly transcendental integral. 
Based on the experience from 
calculating the fully  lightlike limit~\cite{Almelid:2015jia,Almelid:2016lrq,Gardi:2016ttq}, one expects that this would be a property of the specific integral entering the anomalous dimension, i.e. ${\cal T}_{\lambda_i,\lambda_j,\lambda_k}\left[{\cal Y}_{(IJ;KQ)}\right]$.
This is indeed so. Moreover,  each region of this integral, characterized by distinct scaling 
$\lambda_{i}^{n_i\epsilon}\lambda_{j}^{n_j\epsilon}
\lambda_{j}^{n_j\epsilon}$, 
is individually uniformly transcendental. Based on this, using the package {\tt Initial}~\cite{Dlapa:2020cwj}, we obtained a canonical form of the differential equations of each region. The general solution is then obtained, order-by-order in $\epsilon$ in terms of generalized polylogarithms using {\tt PolyLogTools} \cite{Duhr:2019tlz}. 

Important simplifications occur in~$W_{1111}$ upon summing all region integrals. The result is compatible with a  range of physical constraints, including Bose-symmetry, the first-entry condition, a single UV $1/\epsilon$ pole, the known $\beta_Q^2\to 0$ limit obtained in Ref.~\cite{Almelid:2015jia,Almelid:2016lrq,Gardi:2016ttq}, and the known collinear limit for the soft anomalous dimension,  determined by the strict collinear factorization requirement~\cite{Liu:2022elt,Duhr:2025cye}.   These constraints, put together, fully fix the boundary data, yielding a unique result for $W_{1111}$. We stress that in contrast to individual region integrals, this result is free of higher-order poles in~$\epsilon$, and the leading order ${\cal Y}^{(-1)}_{(ij;kQ)}$ is a uniform weight-five MPL function of the variables~(\ref{newVar}) which depends of the expansion parameters (\ref{virtualities_expansion}) only through logarithms. It is also free of singularities at $\textcolor{blue}{\Delta_1}$ and~$\textcolor{blue}{\Delta_2}$. 

The remaining webs in Eq.~\eqref{FourLineWebs}, do not involve functions of all three kinematic variables and are therefore simpler. 
The most complicated of these is $W_{1112}$ (see Fig.~\ref{fig:threeloop4linewebs}), which has one 3-gluon vertex. The calculation strategy is similar to that of the connected webs and will be described in Ref.~\cite{GZ-TBP}. The multi-gluon-exchange webs in Eq.~\eqref{FourLineWebs}, $W_{1122}$ and~$W_{1113}$ (see Fig.~\ref{fig:threeloop4linewebs}), have been computed with all four velocities timelike in terms of MPLs~\cite{Gardi:2013saa,Falcioni:2014pka}. It is therefore straightforward to expand these results according to Eq.~(\ref{virtualities_expansion}).

In addition to the four-line webs in Eq.~\eqref{FourLineWebs}, three-~and two-line webs also contribute to $F_{1,3}$ upon applying colour conservation~\cite{Almelid:2015jia,Almelid:2016lrq}. These contributions can be obtained from known results~\cite{Almelid:2016lrq} or from collinear limits. In the sum of the four-line webs and those with fewer lines, we find that the $\lambda$ dependence entirely cancels out, so the soft anomalous dimension depends only on the three rescaling invariant variables, as expected. The result will be described in the next section. 

\section{The one-mass soft anomalous dimension}
\label{sec:OneMassSoftAD}

Let us summarize the new result. The soft anomalous dimension with one heavy quark and any number of massless particles to three loops takes the form: 
\begin{eqnarray}
    \label{cstructure}
{\mathbf{\Gamma}}
=\mathbf{\Gamma}_{\text{Dip}}+
\mathbf{\Gamma}_{\text{Dip}}^Q +
\mathbf{\Delta}\left(\left\{\rho_{ijkl}\right\}\right)
+
\mathbf{\Delta}^Q\left(\left\{r_{uvQ}\right\}\right)
\,\, +\,\, {\cal O}(\alpha_s^4)\,,
\end{eqnarray}
where the all-massless contributions $\mathbf{\Gamma}_{\text{Dip}}$ and $\mathbf{\Delta}$ have been given in Eqs.~(\ref{Dipole_formula}) and~(\ref{Delta_massless}) respectively, while their counterparts involving the heavy quark $Q$ are:
\begin{equation}
\mathbf{\Gamma}_{\text{Dip}}^Q =
\sum_{i}\mathbf{T}_{i}\cdot\mathbf{T}_{Q}\gamma_{\text{cusp}}(\alpha_s)\log\left(-\frac{\tau\sqrt{p_Q^2}}{2p_i\cdot p_Q}\right)\, +\, \Omega_Q(\alpha_s)\,,
\end{equation}
where the colour-singlet constant term $\Omega_{Q}$ is known to three loops~\cite{Grozin:2014hna,Grozin:2015kna,Bruser:2019yjk} and the multileg contribution is given by 
\begin{align}
    \mathbf{\Delta}^Q\left(\left\{r_{uvQ}\right\}\right)=
\sum_{i<j<k}\sum_{(u,v;w)\in P}
\mathbf{T}_{uv;w Q} {\cal F}_{1,3}(r_{uw Q},r_{vw Q},r_{uvQ})
+ 
\sum_{i<j}\mathbf{T}_{iQ;jQ}{\cal F}_{1,2}(r_{ijQ})\,,
\end{align}
where the three-loop function ${\cal F}_{1,2}^{(3)}$ was computed in Ref.~\cite{Liu:2022elt} and ${\cal F}_{1,3}^{(3)}$ in Ref.~\cite{Gardi:2025lws}. To present the result for  ${\cal F}_{1,3}^{(3)}$ we define, in analogy with 
${\cal F}_{0,4}^{(3)}$ in Eq.~(\ref{Fmassless}),
\begin{align}
\label{calF_to_F}
\begin{split}
    {\cal F}_{1,3}^{(3)}(r_{ijQ},r_{jkQ},r_{ikQ})=
    16\left[F_{1,3}(x; 1-\bar{z},1-z)-F_{1,3}\left(x; z,\bar{z}\right)\right],
    \end{split}
\end{align}
where the variables $\{x,z,\bar{z}\}$ are defined by 
\begin{align}
\label{newPara}
\begin{split} r_{ijQ}\equiv\,\,&
\frac{\beta_Q^2\beta_i\cdot\beta_j}{2\beta_i\cdot\beta_Q \beta_j\cdot\beta_Q}
=\frac{x(x+z-\bar{z})}{(1-z)(1-\bar{z})},
   \\
    r_{jkQ}\equiv\,\,&
\frac{\beta_Q^2\beta_j\cdot\beta_k}{2\beta_j\cdot\beta_Q \beta_k\cdot\beta_Q}  =
    \frac{x(x+z-\bar{z})}{z\bar{z}},
    \\
    r_{ikQ}\equiv\,\,&
\frac{\beta_Q^2\beta_i\cdot\beta_k}{2\beta_i\cdot\beta_Q \beta_k\cdot\beta_Q}
    =
    \frac{x(x+z-\bar{z})}{(1-z)(1-\bar{z})z\bar{z}}\,,
\end{split}
\end{align}
allowing us to express $F_{1,3}$ using MPLs without square roots. In~\cite{MathematicaNotebook} we provide the explicit expression of $F_{1,3}$, written in terms of uniform weight-five MPLs of $\{x,z,\bar{z}\}$ and Riemann zeta values with rational numerical coefficients. 
Permutation symmetry in $i\leftrightarrow k$ is realised in Eq.~(\ref{calF_to_F}) via $z\leftrightarrow 1-\bar{z}$ antisymmetry, just as in Eq.~(\ref{Fmassless}).
Note that $F_{1,3}$ is invariant under the following Galois symmetries,
\begin{align}
\label{GaloisSymm}
\begin{split}
F_{1,3}(x; z,\bar{z})=\,\,&F_{1,3}(-x; \bar{z},z)= F_{1,3}(-x-z+\bar{z}; z,\bar{z}),
    \end{split}
\end{align}
related to flipping the sign of the square roots that appear upon solving~\eqref{newPara} for $\{x,z,\bar{z}\}$. The full one-mass quadrupole in~Eq.~(\ref{cstructure}) is obtained by performing the three permutations on $F_{1,3}$ of Eq.~(\ref{calF_to_F}), as prescribed by~$P$ below Eq.~(\ref{Delta_massless}).

In the new variables, the alphabet of ${\cal F}_{1,3}^{(3)}$ is 
\begin{align}
\label{alphabet}
\begin{split}
    &\{\omega_i\}^{\text{phy}}=\,\,\{z,\bar{z},1-z,1-\bar{z}\}
    \cup\{x,\bar{z}-z-x\}\cup\{x+z,
   \\&\hspace{20mm}\,\, \bar{z}-x,
 1-z-x,1-\bar{z}+x,z+x-z\bar{z},\bar{z}-x-z\bar{z}\}\,.
\end{split}
\end{align}
These 12 letters represents physical collinear singularities,  $\beta_u\cdot \beta_v\to 0$. The first group of four letters corresponds
to $r_{uvQ}\to \infty$ and coincides with the full alphabet of~${\cal F}_{0,4}^{(3)}$ in  Eq.~(\ref{Fmassless}). 
The next group of two letters corresponds to $r_{uvQ}\to 0$, while the final group of six letters to $r_{uvQ}\to 1$. It is convenient to interpret these limits in the $\beta_Q$ rest frame. There 
$ r_{uv Q}=\frac{1}{2}\left(1-\cos\theta_{uv}\right),
$
where $\theta_{uv}$ is the spatial angle between particles  $u$ and~$v$. The singularities at $r_{uv Q}\to \{0,1\}$ then correspond the collinear ($\theta_{uv}\rightarrow 0$) and the back-to-back ($\theta_{uv}\rightarrow \pi$) configurations, respectively. The first symbol entry of ${\cal F}_{1,3}^{(3)}$, associated with its channel discontinuities, is more restricted and only consists of the three ratios $r_{uvQ}$, with branch cuts for $r_{uvQ}>0$.
\begin{figure}[t]
\centering \includegraphics[width=1\linewidth]{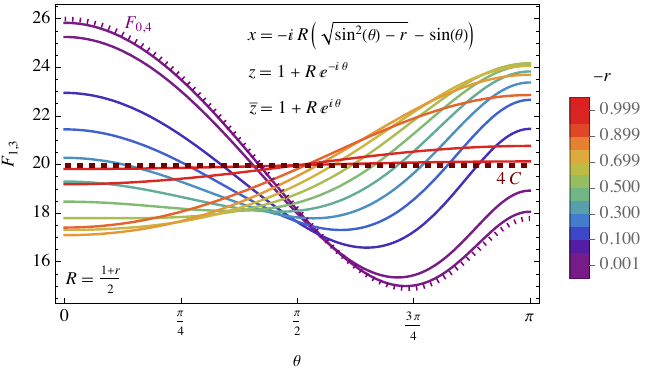}
    \caption{$F_{1,3}(x; z,\bar{z})$ in Euclidean kinematics with $(z,\bar{z})$ parametrized in polar coordinates about the point $z=\bar{z}=1$,
    sampled along the line $R=(1+r)/2$ at fixed values of $r=r_{ijQ}<0$, interpolating between the collinear limit $\beta_j||\beta_k$ ($R=0$) of Eq.~(\ref{collF})  and the lightlike limit of Eq.~(\ref{lightlike_limit}) at $r=0$. \label{fig:F13Euclid} }
\end{figure}

The collinear limits of ${\cal F}_{1,3}^{(3)}$ are consistent with the strict collinear factorization constraints~\cite{Liu:2022elt,Duhr:2025cye}, yielding 
\begin{align}
\label{coll}
     \begin{split}
   {\cal F}_{1,3}^{(3)}(0,r,r)=-{\cal F}_{1,3}^{(3)}(r,0,r)= 2{\cal F}_{0,3}^{(3)},
     \end{split}
     \end{align}
or equivalently, for $F_{1,3}(x;z,\bar{z})$,
\begin{align}
\label{collF}
\begin{split}
    F_{1,3}(0;1,1)
    =4C\,,
    \quad
     F_{1,3}(0;0,0)
     =0\,.
     \end{split}
 \end{align}
The lightcone  limit $\beta_Q^2\to 0$~is obtained by taking $x\to 0$,  
\begin{equation}
\label{lightlike_limit}
F_{1,3}(0;z,\bar{z})=F_{0,4}(z,\bar{z})\,,
\end{equation}
recovering~the known result~\cite{Almelid:2015jia} of~Eq.~(\ref{Fmassless}). Eq.~(\ref{lightlike_limit}) also represents the triple collinear limit, $\beta_i||\beta_j||\beta_k$~\cite{Duhr:2025cye,Gardi:2025ule}. 
Fig.~\ref{fig:F13Euclid} shows $F_{1,3}(x; z,\bar{z})$ in the Euclidean domain, where it is real-valued. The figure illustrates that the computed result satisfies both the two-particle collinear limit (\ref{collF}) and the massless limit~(\ref{lightlike_limit}).

\section{Colliner limits}

Collinear limits are the most straightforward limits to consider, when specialising the wide-angle scattering kinematics. It is well known that an $n$ parton amplitude ${\cal M}_n$ factorises in the collinear limit where massless partons 1 and 2 become collinear, into a splitting amplitude times an $(n-1)$ parton amplitude~\cite{Kosower:1999xi,Catani:2011st,Feige:2014wja}. If both partons 1 and 2 belong to the final state, then the factorization is \emph{strict}, in the sense that the splitting amplitude depends only on the degrees of freedom of the collinear partons -- not on the rest of the process. The factorization structure is given by  
\begin{align}
    {\cal M}_n(p_1,p_2,\{p_j\};\mu)\xrightarrow{\;{1\parallel 2}\;}\textbf{Sp}(p_1,p_2;\mu){\cal M}_{n-1}(p_1+p_2;\{p_j\};\mu)\,,
\end{align}
where $\{p_j\}$ represents the rest of the process, the non-collinear partons $j=3,\ldots, n$.
This remains valid for IR singularities: the IR singularities of the splitting amplitude, governed by $\mathbf{\Gamma}_{\textbf{Sp}}$,  are the singularities of the $n$ parton amplitude which are not singularities of the $n-1$ amplitude. In terms of the anomalous dimension, this corresponds to the following relation~\cite{Becher:2009qa}
\begin{align}
\label{Gamma_sp}
  \mathbf{\Gamma}_{\textbf{Sp}}=\mathbf{\Gamma}_n-\mathbf{\Gamma}_{n-1}\,.
\end{align}
This relation translates into a constraint on the soft anomalous dimension $\mathbf{\Gamma}_n$ itself, upon requiring \emph{strict} collinear factorization, that is requiring that $\mathbf{\Gamma}_{\textbf{Sp}}$ only depends on the colour and kinematic degrees of freedom of the two particles becoming collinear; the dependence on the rest of the process must cancel out on the right hand-side of Eq.~(\ref{Gamma_sp}).
It is straightforward to show that Eq.~(\ref{Gamma_sp}) is readily realised for the dipole formula (\ref{Dipole_formula}) upon using colour conservation. It provides valuable constraints on corrections to the dipole formula at any loop orders. 
The consequences regarding the three- and four-loop soft anomalous dimension for massless scattering have been studied 
in Refs.~\cite{Becher:2009qa,Almelid:2017qju,Maher:2023jqy,Duhr:2025cye}. Ref.~\cite{Duhr:2025cye} focused on extending the analysis to multi-collinear limits, where for $m$ particles getting simultaneously collinear 
\begin{align}
\mathbf{\Gamma}_{\textbf{Sp}}^{(m)}=\mathbf{\Gamma}_n-\mathbf{\Gamma}_{n-m+1}\,,
\end{align}
the strict collinear factorization requirement is again that $\mathbf{\Gamma}_{\textbf{Sp}}^{(m)}$ would only depend on the degrees of freedom of the $m$ collinear partons.
Ref.~\cite{Duhr:2025cye} established that for purely massless scattering, imposing the strict collinear factorization constraints in all two-particle collinear limits, implies that strict collinear factorization also holds in all multi-collinear limits. Interestingly, when also heavy quarks are present, new constraints appear.

The one-mass soft anomalous dimension discussed in the previous section admits a plethora of constraints due to strict collinear factorization, most of which were predicted in Ref.~\cite{Liu:2022elt}.  Specifically, as stated in Eq.~(\ref{coll}) above, two-particle collinear limits relate the kinematically-dependent one-mass functions ${\cal F}_{1,3}(r_{uw Q},r_{vw Q},r_{uvQ})$ and ${\cal F}_{1,2}(r_{ijQ})$ to the same weight-5 transcendental constant ${\cal F}^{(3)}_{0,3}=32 C$ appearing in the all-massless soft anomalous dimension, Eq.~(\ref{F03massless}) above. Note that these constraints coincide with those admitted by all-massless function ${\cal F}^{(3)}_{0,4}$. 

Furthermore, there is an interesting constraint on three-particle collinear limit from strict collinear factorization, which has been discussed 
in Refs.~\cite{Duhr:2025cye,Gardi:2025lws,GZ-TBP}. We have seen that $F_{1,3}$ depends on only the three rescaling invariant cross ratios defined in Eq.~\eqref{newVar}.
As a consequence of the  rescaling symmetry with respect to $\beta_Q$, there are two physically-distinct limits of the velocities,  which correspond to the very same limit of the cross ratios in Eq.~(\ref{newVar}), namely \hbox{$r_{ijQ}\sim r_{ikQ}\sim r_{jkQ}\sim\lambda$}, with the ratios between any two of them remaining finite for $\lambda\to 0$. The two limits are the massless limit:
\begin{align}
\beta_Q^2\sim \lambda,
\qquad 
\beta_i\cdot \beta_j\sim 
\beta_i\cdot \beta_k\sim 
\beta_j\cdot \beta_k\sim {\cal O}(\lambda^0)\,,
\end{align}
and the triple collinear limit:
\begin{align}
    \beta_Q^2\sim {\cal O}(\lambda^0),
\qquad 
\beta_i\cdot \beta_j\sim 
\beta_i\cdot \beta_k\sim 
\beta_j\cdot \beta_k\sim \lambda\,,
\end{align} 
where in either of these cases, one considers 
\begin{align}
\beta_i\cdot \beta_Q\sim 
\beta_j\cdot \beta_Q\sim 
\beta_k\cdot \beta_Q\sim 
{\cal O}(\lambda^0)\,.
\end{align}
These two limits represent, respectively, the lightcone expansion of particle $Q$, and the ``complementary'' lightcone expansion, where all other particles become collinear. We stress that this is true in the presence of an arbitrary momentum recoil, so this is a non-trivial relation two between physically distinct limits of the anomalous dimension,  which coincide because of the rescaling symmetry. This implies that for function $F_{1,3}$, the constraint from the triple collinear limit is the same as Eq.~\eqref{lightlike_limit}. 
Ref.~\cite{Gardi:2025ule} explained that it is a general feature, stemming from rescaling symmetry, that distinct physical limits are described by the same mathematical limit. This has the potential of identifying additional non-trivial relations in QCD. 

\section{Conclusions}

We reviewed recent progress on the infrared singularities of QCD amplitudes. The soft anomalous dimension is rich and yet remarkably simple, especially when compared to finite corrections to multileg amplitude at similar loop orders. 
The three-loop soft anomalous dimension for massless scattering has been computed a decade ago, while the computation of the fully-massive one is still beyond reach with present methods. 
The primary step forward has been the computation of the three-loop soft anomalous dimension for one massive particle with any number of massless ones. It was made possible by a new strategy that fully exploits the simplification granted by the  lightlike limit using the Method of Regions.
This strategy can be further used to determine the three-loop soft anomalous dimension for amplitudes involving two massive particles, a top quark pair, and potentially compute the soft anomalous dimension for massless scattering at four loops. 

Beyond explicit computations, the study of the soft anomalous dimension provides insight into multileg amplitudes at high orders in the loop expansion, and a laboratory to explore these. This research connects with the study of special kinematic limits such as (multi-) collinear limits~\cite{Becher:2009qa,Almelid:2017qju,Duhr:2025cye} and the (multi-) Regge limit~\cite{Korchemskaya:1994qp,Korchemskaya:1996je,DelDuca:2011wkl,DelDuca:2011ae,DelDuca:2014cya,Caron-Huot:2017fxr,Caron-Huot:2017zfo,Caron-Huot:2020grv,Falcioni:2020lvv,Falcioni:2021buo,Falcioni:2021dgr,Caola:2021izf,Buccioni:2024gzo}, where new factorization properties arise. We have repeatedly seen that these limits provide boundary data as well as important checks of the computation of soft singularities in general kinematics. 
The analytic structure of the soft anomalous dimension can also be explored and further constrained using cut techniques and differential equations. The knowledge gained by these considerations,  alongside special kinematic limits, is valuable in its own right, but it may also open alternative avenues to determine the soft anomalous dimension and related quantities via a bootstrap approach, continuing the work of Ref.~\cite{Almelid:2017qju}. 

\acknowledgments

This talk was given at RadCor 2025, the 17th International Symposium on Radiative Corrections: Applications of Quantum Field Theory to Phenomenology in Puri, India. 
ZZ has been supported by the China Scholarship Council PhD programme. 
EG is supported by the STFC Consolidated Grant \emph{Particle Physics at the Higgs Centre}. 
For the purpose of open access, the authors have applied a Creative Commons Attribution (CC BY) licence to any Author Accepted Manuscript version arising from this submission.

\bibliographystyle{JHEP}
\bibliography{biblio}

\end{document}